\documentclass[prl,showpacs,aps,twocolumn,nofootinbib,superscriptaddress]{revtex4}
\usepackage{graphicx,amssymb,amsmath}

\begin{document}

\title{
Interplay between phase defects and spin polarization
in the specific heat of \\
the spin density wave compound (TMTTF)$_2$Br in a magnetic
field
}

\author{R. M\'elin}
\affiliation{
Centre de Recherches sur les Tr\`{e}s Basses Temp\'{e}ratures, 
CNRS, BP 166, 38042 Grenoble cedex 9, France}
\author{J.C. Lasjaunias}
\affiliation{
Centre de Recherches sur les Tr\`{e}s Basses Temp\'{e}ratures, 
CNRS, BP 166, 38042 Grenoble cedex 9, France}
\author{S. Sahling}
\affiliation{
Centre de Recherches sur les Tr\`{e}s Basses Temp\'{e}ratures, 
CNRS, BP 166, 38042 Grenoble cedex 9, France}
\affiliation{Institut f\"ur Festkorperphysik, IFP,
Technische Universit\"at Dresden, D-01069 Dresden, Germany}
\author{G. Remenyi}
\affiliation{
Centre de Recherches sur les Tr\`{e}s Basses Temp\'{e}ratures, 
CNRS, BP 166, 38042 Grenoble cedex 9, France}
\author{K. Biljakovi\'c}
\affiliation{
Centre de Recherches sur les Tr\`{e}s Basses Temp\'{e}ratures, 
CNRS, BP 166, 38042 Grenoble cedex 9, France}
\affiliation{Institute of Physics,
P.O. Box 304, Zagreb, Croatia}

\begin{abstract}
Equilibrium heat relaxation experiments provide
evidence that the ground state of the commensurate spin
density wave (SDW) compound (TMTTF)$_2$Br after
the application of a sufficient magnetic field is different from the
conventional ground state. The experiments are interpreted 
on the basis of the local model of strong pinning as 
the deconfinement of soliton-antisoliton pairs 
triggered by the Zeeman coupling to
spin degrees of freedom, resulting in a
magnetic field induced
density wave glass for the spin carrying phase configuration.
\end{abstract}

\date{\today}
\maketitle

Metastability results from energy minima in a system phase space,
separated from each other by energy barriers. In some complex systems
such as spin glasses, the number of
metastable states and the scaling of the energy barriers with the
system size are such that ergodicity is broken, meaning that
time averaging is not equivalent to ensemble averaging
because the system is ``trapped'' in a valley of the energy landscape.
Metastability is also found in model systems
such as molecular magnets or Josephson junctions, 
described by an energy potential $V(\varphi)$
as a function of a single degree of freedom $\varphi$.
We investigate below experimentally and theoretically 
the residual degrees of freedom of
a spin density wave (SDW) at very low temperature in a magnetic field,
interpreted as
the properties of a classical potential
$V(\varphi(y_i))$ for the SDW phase $\varphi(y_i)$ 
at the coordinate $y_i$ along the chain 
of a strong pinning impurity.

Below the Peierls transition temperature, 
charge density waves (CDWs) and SDWs
in quasi-one-dimensional (quasi-1D) compounds
are characterized by a spatial modulation of the electronic density
(or spin) along the chains. The phase profile is the result of
a compromise between the elastic energy that penalizes
large phase gradients, and the pinning energy that
tends to fix the phase at the strong pinning centers.
These two ingredients of the Fukuyama-Lee-Rice model \cite{FLR}
lead to metastability as the result of collective pinning.
Collective pinning is however frozen below
a glass transition of order $\sim 50$~K, as shown by
dielectric susceptibility experiments \cite{Damir}. At very low
temperature, 
the residual degrees of freedom
in a zero magnetic field
correspond to
the local defects of the local model of strong pinning
\cite{Fuku,Larkin,Ov,Melin02,revue,Abe}.
Larkin \cite{Larkin} and Ovchinnikov \cite{Ov}
have shown that a single strong pinning
impurity leads to
a bound state of an electron-like soliton and a
hole-like antisoliton. This results in a potential $V(\varphi(y_i))$
with multiple minima \cite{Melin02}, leading to slow relaxation
in agreement
with the very low temperature heat relaxation experiments \cite{Ov}.

In order to explore the role of a magnetic field,
we choose the compound (TMTTF)$_2$Br
with a sufficiently narrow spectrum of relaxation times
because of its commensurate (antiferromagnetic) ground state
\cite{ECRYS02,Lasjau05}. This allows
a systematic study of the {\it equilibrium}
energy relaxation over almost one decade in temperature
and a direct comparison to the local model of strong
pinning without introducing an additional time
scale related to ageing \cite{Melin02,Biljak91}.

\begin{figure}[tbp]
\centerline{\includegraphics[width=7cm]{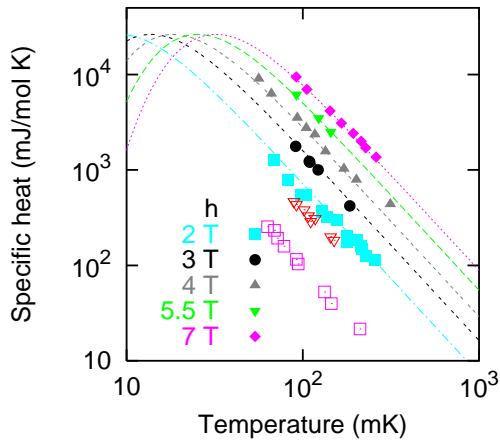}}
\caption{(Color online.) Temperature dependence of the specific heat
for various magnetic fields.
($\Box$, purple)
corresponds to $h=0$~T in
a zero field cooled sample, and ($\triangledown$, red)
corresponds to $h=0$~T in a zero field cooled sample where the magnetic field
was cycled according to $h=0 \mbox{ T} \rightarrow 5.5 \mbox{ T}
\rightarrow 0 \mbox{ T}$.
The solid lines are a fit to $C_p^{(0)}(\mu h/k_B T)$.
}
\label{fig1}
\end{figure}

\begin{figure}[tbp]
\centerline{\includegraphics[width=7cm]{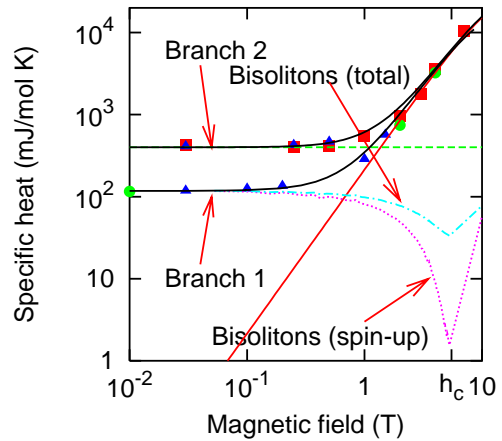}}
\caption{(Color online.) Magnetic field dependence of the specific heat at
a fixed temperature $T=92$~mK.
The magnetic field is first cycled from $h=0$~T
to $h=7$~T, starting from a zero
field cooled sample ($\bullet$, green, branch 1),
and next decreased back
from $h=7$~T to $h=0$~T ($\blacksquare$, red, branch 2). 
The system remains on branch 2 after additional magnetic
field cycles. Recovery branch $1$ starting from branch $2$
($\blacktriangle$) is explained in the text .
The solid (red) line ``1'' is a fit to $C_p^{(0)}(\mu h/k_B T)$
(parameters as on Fig.~\ref{fig1}).
Branch 2 saturates to $C_0^{(2)}$ for $h\ll h_c$.
The spin-up bisoliton contribution (dotted lines) vanishes at $h\sim h_c$.
}
\label{fig2}
\end{figure}

The specific heat of about $60$~mg of
a (TMTTF)$_2$Br salt was measured at CRTBT-Grenoble in a dilution
cryostat set-up under magnetic field up to $7$~T. This compound was
previously investigated in zero field in a similar
temperature range \cite{ECRYS02}. We use a standard relaxation
method \cite{ECRYS02,Lasjau05},
but in contrast to previous experiments, the specific heat is determined
at equilibrium, once the heat relaxation regime 
$\Delta T(t)$ as a function of time $t$,
obtained after a long heat input, becomes exponential.
This internal relaxation time $t_{in}$ reaches up to more than $10^4$~s at
the temperature
$T=60$~mK and at high field.
This technique requires probing times of
about $3\div4 \, t_{in}$ for the determination of the
time constant $\tau_{\rm eq}$ of the exponential
relaxation for $t>t_{in}$, and hence extremely
stable field supply and temperature regulation. The equilibrium specific
heat $C_{eq}$
reported on Figs.~\ref{fig1} and~\ref{fig2} is obtained from the
relation $C_{eq}=\tau_{eq}/R_{hl}$, where $R_{hl}$ is the heat link
resistance between the sample and the regulated cold sink, measured
at equilibrium under permanent power supply (more details will
be published separately).

The temperature dependence of the specific heat $C_p(T)$ is shown on
Fig.~\ref{fig1} for different values of the magnetic field.
We find experimentally a specific heat proportional to $1/T^2$,
with a prefactor
varying by almost two orders of magnitude when the magnetic
field increases from $h=0$~T to $h=7$~T. 
This temperature variation
is interpreted \cite{Larkin,Ov}
as the high temperature tail of a Schottky
due to two level-like systems.
The specific heat of a concentration $A$ of
two-level systems with an energy splitting $\Delta$ is
$C_p^{(0)}(\Delta/k_B T)=A (\Delta/k_B T)^2/[2
\cosh{(\Delta/2k_B T)}]^2
$, that  behaves like
$C_p^{(0)}(\Delta/k_B T)\simeq A (\Delta/2 k_B T)^2$ for
$k_B T$ large compared to $\Delta$.
The temperature dependence of the specific heat
is well described
by $C_p^{(0)}(\mu h/k_B T)$, with a splitting $\Delta=\mu h$.
The value of $\mu\simeq 0.011 \mu_B$
(with $\mu_B$ is the Bohr magneton
of an electron)
deduced from the fit on Fig.~\ref{fig1}
is approximately one order of magnitude smaller
than the SDW amplitude measured by NMR \cite{NMR}.
This is because solitonic excitations involve a distribution
of spins with a staggered orientation, with a net magnetic moment
smaller than the moment of an individual spin (see Fig.~\ref{fig4}).
We have chosen the fit leading to the
smallest value of $A\simeq 60$~J/mol K
compatible with experiments. 
This results in a huge number of defects induced by the magnetic field
(approximately $8$ defects per unit cell) that cannot be explained by
impurities \cite{JLTP03}.
Here we relate the magnetic field specific heat
to spin degrees of freedom in a density wave (DW) glass,
not to the phase excitations of bisolitons in the local
model of strong pinning where the number of phase defect would be
equal to the number of pinning centers.
The fit of the temperature dependence of the
specific heat in zero field leads to a concentration of defects of
approximately $20\,\%$. This concentration is consistent
with the Ovchinnikov estimate in (TMTSF)$_2$PF$_6$ \cite{Ov},
but is however too large to be ascribed to extrinsic impurities.

We report for the first time metastability induced by the magnetic field
at a fixed temperature $T=92\,$mK (see Fig.~\ref{fig2}).
The sample that has not ``seen'' the magnetic field follows
the branch $C_p^{(1)}(h)$
and remains on branch 1 if the applied magnetic field
does not exceed $h_c\sim 5.5$~T. If the magnetic field increases
above $h_c$, the specific heat follows 
$C_p^{(2)}(h)$ upon decreasing the
magnetic field below $h_c$, and remains on branch 2
if the magnetic field is further cycled above $h_c$.
This signals that a magnetic field $h>h_c$ induces a new ground state
that we identify below as a density wave glass.
Branch $1$ is recovered with the following history:
i) the sample is reheated in zero field from
$90$~mK up to above $20$~K ii) once the sample is cooled down,
the magnetic field is increased up to $1.5$~T and decreased again
to $0.2$~T.
The raw experimental data are well fitted by
$C_p^{(1,2)}(h,T)=\left[C_0^{(1,2)}+ C_p^{(0)}(\mu h/k_B T)
\right]$~mJ/mol K for branches $1$ and $2$,
where $h$ is in Tesla and $T$ in Kelvin, and with
$C_0^{(1)}=120$~mJ/mol K, and
$C_0^{(2)}=404$~mJ/mol K.
We find $C_p^{(0)}(\mu h/k_B T)\simeq
190 (h/T)^2$~mJ/mol K for $k_B T$ much larger
than $\mu h$.

\begin{figure}[tbp]
\centerline{\includegraphics[width=7cm]{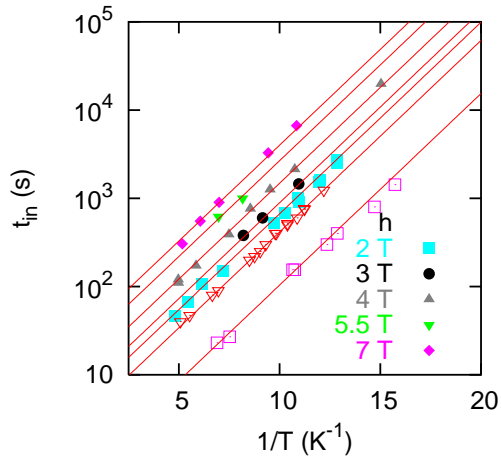}}
\caption{(Color online.) Ahrrenius plot $t_{in}$
{\it versus} $1/T$ of the internal relaxation time.
The symbols correspond to the same values of the magnetic field
as on Fig.~\ref{fig1}. The experimental data are fitted
by $t_{in} \simeq \tau_0(h) \exp{(E_A/T)}$, with
$E_A=0.5$~K and $\tau_0(2 \, T)=4.5$~s,
$\tau_0(3 \, T)=6$~s,
$\tau_0(4 \, T)=10$~s,
$\tau_0(5.5 \, T)=18$~s,
$\tau_0(7 \, T)=27$~s,
$\tau_0(0 \, T)=0.7$~s (branch 1),
$\tau_0(0 \, T)=2.8$~s (branch 2).
}
\label{fig3}
\end{figure}

\begin{figure}[tbp]
\centerline{\includegraphics[width=6cm]{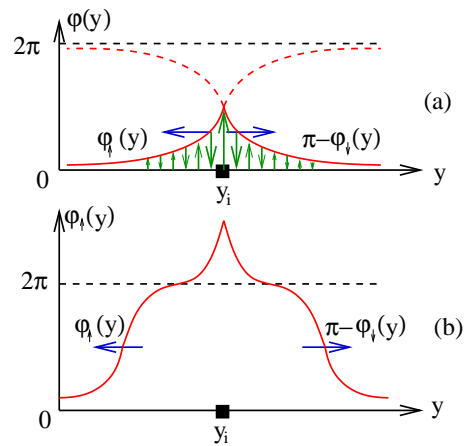}}
\caption{(Color online.) (a): A spin polarized bisoliton
made of the superposition of a spin-up electron-like soliton
in the field $\varphi_\uparrow(y)$
and a spin-down hole-like antisoliton in the field
$\varphi_\downarrow(y)$ is deconfining (according to the arrows).
The dashed lines are the continuation of the two solutions,
without physical meaning.
(b): The solitons and the antisolitons
are deconfined, and a new bisoliton is generated.
In the spin polarized case,
the energy of (b) is equal to to the energy of (a) minus
the spin energy
$4\pi h \rho_0 Q^{-1}$. The impurity is indicated by the black box
at position $y_i$ along the chain.
}
\label{fig4}
\end{figure}

The internal equilibrium time $t_{in}$ follows an activated regime
$t_{in}=\tau_0 \exp{(E_A/T)}$ with an activation energy
$E_A\simeq 0.50$~K and a magnetic field-dependent attempt time
$\tau_0$ of the order of a few seconds (see Fig.~\ref{fig3}).
In zero field,
we find an excellent agreement
with a previous determination of the activation energy,
obtained from the spectrum of relaxation times
\cite{ECRYS02}. The activated behavior is in agreement with
the local model of strong pinning that we consider now.

The effective
1D Hamiltonian for the phase $\varphi(y)$
of the SDW takes the form
\cite{FLR,Fuku,Larkin,Ov,Melin02,revue,Abe}:
\begin{equation}
\label{eq:H-phase}
{\cal H} = \frac{v_F}{4\pi} \int dy 
\left(\frac{\partial \varphi(y)}{\partial y}
\right)^2 +
w \int dy 
\left[1-\cos{\varphi(y)}\right]
,
\end{equation}
where $y$ is the coordinate along the chain, $v_F$ the
Fermi velocity, $w$ the commensuration potential.
The excitations in the absence of 
impurities are pairs of $\pm 2 \pi$
solitons and antisolitons,
the phase of which winds by $\pm 2 \pi$
within a correlation length $\xi=\sqrt{\hbar v_F/2\pi w}$.
A SDW is viewed as the superposition of
two out-of-phase CDWs for spin-up and spin-down electrons, and
the SDW pinning energy is obtained to second order as
\cite{Tua,Suzumura,Tutto,Maki}
${\cal H}_{\rm imp} = \sum_{i}
V_i  \cos{\left[2\left(Q y_i+\varphi(y_i)\right)\right]}$,
where the sum runs over all strong pinning impurities at
position $y_i$ along the chain.
The phase field is $\varphi(y)=\varphi_\uparrow(y)
-\varphi_\downarrow(y)+\pi$, where $\varphi_\uparrow(y)$
and $\varphi_\downarrow(y)$ are the spin-up and spin-down phase fields.
The residual specific heat at low field 
of $120$~mJ/mol K in $C_p^{(1)}(h,T)$ 
is related to metastability due to
spinless bisolitons (bound state of a
soliton and an antisoliton) \cite{Larkin,Ov}.
The phase profile $\varphi_b(y)$ of a bisoliton is \cite{Larkin}
\begin{equation}
\label{eq:phib}
\tan{\left(\frac{\varphi_b(y)}{4}\right)}
=\tan{\left(\frac{\psi_i}{4}\right)}
\exp{\left(-\frac{|y-y_i|}{\xi}\right)}
.
\end{equation}
The ground state of a bisoliton in the absence of a magnetic field
is at energy $E_0$,
separated by a barrier from a metastable state 
at energy $E_0+\Delta E$, with
$\Delta E = 4 E_S = 16 w \xi$, where $E_S$ is the energy
of a $\pm 2 \pi$ soliton in the pure system.
This defines an effective two-level system \cite{Larkin,Ov}.

A Zeeman coupling to the magnetic field is included
now through
${\cal H}_{h}=-\mu h \int dy \left[\rho_\uparrow(y)-
\rho_\downarrow(y)\right]$,
where the electronic density $\rho_\sigma(y)$,
with $\sigma=\uparrow,\downarrow$, is defined by \cite{revue}
\begin{equation}
\label{eq:rho}
\rho_\sigma(y)=\rho_0 \left[ 1 + Q^{-1}
\frac{\partial \varphi_\sigma(y)}{\partial y} \right]
+\rho_1 \cos{(Q y + \varphi_\sigma(y))}
,
\end{equation}
where $Q$ is the SDW wave-vector,
$\rho_0$ is the electronic charge per unit length in the absence of
deformation.
The term containing $\rho_1$ is relevant to local pinning,
but averages to zero in the
Zeeman energy integrated over $y$
because of the short scale oscillations at the
Fermi wavelength.

A $2\pi$ soliton in
the spin-up field $\varphi_\uparrow(y)$
and a $-2\pi$ soliton in the spin-down field
$\varphi_\downarrow(y)$
both carry a net spin-up because their energy decreases
by $2\pi \mu h \rho_0 Q^{-1}$ in a magnetic field $h$.
In the absence of impurities,
the SDW ground state is unstable
against the nucleation of pairs of solitons 
and antisolitons carrying a net spin-up
if the gain in the Zeeman energy exceeds the soliton energy
$E_S$, a condition equivalent to $h>h_c$, with
$h_c=2 Q E_S/\pi \mu \rho_0$.
An upper bound to the 
temperature of the maximum of the Schottky anomaly in a zero field is
$T_{\rm max}\simeq 10\div 20$~mK, leading to $4 E_S\simeq2.4 T_{\rm max}$, 
so that $h_c=4 E_S/\mu$ is lower than $\sim 2.2 \div 4.3$~T is compatible with
the cross-over magnetic field obtained experimentally.

\begin{figure}[tbp]
\centerline{\includegraphics[width=9cm]{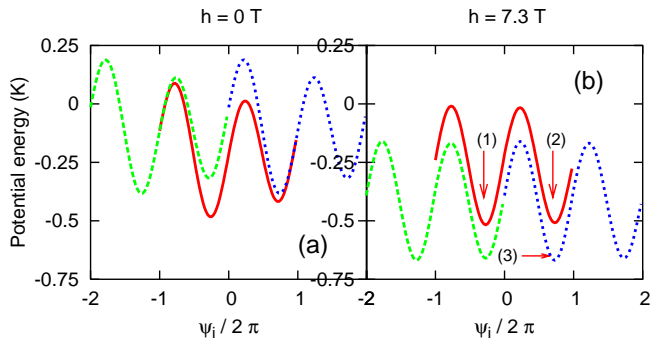}}
\caption{(Color online.) Energy potential (in K) as a function of the
phase at the position $y_i$ of a strong pinning impurity,
(a): for $h=0$, and (b): for $h=7.3$~T (or $\mu h=0.08\,$K).
The pinning energy is
such as to reproduce the experimental activation energy
$E_A \simeq 0.5\,$K
(see Fig.~\ref{fig3}).
The figures show the potential energies of a bisoliton with no additional
soliton/antisoliton pair (solid line, red), with a soliton for
$y<y_i$ and an antisoliton for $y>y_i$
(short dashed, blue), 
and with an antisoliton for $y<y_i$ and a soliton for $y>y_i$
(long dashed, green).
}
\label{fig5}
\end{figure}

Now, we show how deconfined pairs of
solitons and antisolitons are nucleated from the strong
pinning impurities.
A spinless bisoliton corresponds to
$\varphi_\uparrow(y)=\varphi_b(y)$ and 
$\pi-\varphi_\downarrow(y)=0$
for all values of $y$ (see Eq.~(\ref{eq:phib})
for $\varphi_b(y)$).
Alternatively, the field $\varphi_\downarrow(y)$
can be excited:
$\varphi_\uparrow(y)=0$
and 
$\pi-\varphi_\downarrow(y)=\varphi_b(y)$
for all values of $y$.
On the other hand,
a maximally spin polarized bisoliton is obtained by reversing the 
spin in the part of the soliton corresponding to $y>y_i$, in such
a way that
$
\varphi_\uparrow(y)=\varphi_b(y)$ and 
$\pi-\varphi_\downarrow(y)=0$
for $y<y_i$, and
$\varphi_\uparrow(y)=0$ and 
$\pi-\varphi_\downarrow(y)=\varphi_b(y)$
for $y>y_i$.
A sequence generating a deconfined soliton-antisoliton
pair starts from the minimum (1) with a small value of $\psi_i/2\pi$
(see Fig.~\ref{fig5}). The phase $\psi_i$ at
the position of the impurity crossed over to the other minimum (2) with
$\psi_i/2\pi \simeq 1$, and the
soliton-antisoliton pair deconfines by relaxing to the minimum
(3). The process can then be iterated, inducing the same
random configuration of the SDW phase as in a DW glass \cite{Fuku}.
The spin degrees of freedom on
the solitons and antisolitons 
define two-level systems
with a splitting proportional to the applied magnetic field,
which explains the
high-field contribution $C_p^{(0)}(\mu h/k_B T)$
proportional to $(\mu h/k_B T)^2$ (see Fig.~\ref{fig2}).

The bisoliton contribution in low field
(see Fig.~\ref{fig2})
is obtained by assuming a population of
spinless and spin polarized bisolitons in thermal equilibrium.
The specific heat of the spin-up bisolitons with a magnetic
moment parallel to the applied magnetic field vanishes at
the magnetic field $h_c$ (see Fig.~\ref{fig2}). For this value
of the magnetic field,
the magnetic energy is opposite to the Larkin-Ovchinnikov
level splitting in zero field, resulting in a degenerate
two-level system with a vanishingly small specific heat.
The spinless bisolitons
restore a finite specific heat (see the total contribution
of the bisolitons on Fig.~\ref{fig2}).
If the field is reduced from a large value,
the pairs of solitons and antisolitons
annhilate reversibly for $h>h_c$, and
the phase is trapped at $h=h_c$.
The random phase pattern induced by the magnetic field 
persists if the field decreases back to zero.
The spin degrees of freedom on
the residual phase defects explain the specific heat of branch 2,
larger than the specific heat of branch~1.

To conclude, we have found experimentally a very persistent
metastable branch in the magnetic field dependence of the specific
heat with a larger specific heat than in the zero
field case. We interpreted this observation
in terms of the local
model of strong pinning coupled to a Zeeman field, in which
we find an instability of the SDW ground state with a flat phase
towards a DW glass with a random phase configuration.
The additional contribution to the specific heat when
coming back to zero field is explained
by the spin entropy due to the magnetic moments on the
phase defects of the DW glass.
Interestingly, a moderate pressure induces a transition
to an incommensurate phase in the same compound 
\cite{Klemme}, which was explained by the formation
of discommensurations.

The authors thank S. Brazovskii for a useful correspondance on
a preliminary version of this work.
R.M. acknowledges a useful discussion with M. Mueller.
S.S. acknowledges CRTBT for financial support for his
stay, in particular P. Monceau for continuous interest
in this work. CRTBT is associated with Universit\'e Joseph
Fourier.


\begin{thebibliography}{99}
  
\bibitem{FLR} H. Fukuyama and P. A. Lee, Phys. Rev. B
  {\bf 17}, 535 (1977);
  H. Fukuyama, J. Phys. Soc. Jpn.
  {\bf 45}, 1474 (1978);
  P. A. Lee and T. M. Rice, Phys. Rev. B {\bf 19}, 3970 (1979).

\bibitem{Damir} D. Stare\v{s}ini\'c {\it et al.},
Phys. Rev. B {\bf 65}, 165109 (2002).

\bibitem{Fuku} H. Fukuyama, J. Phys. Soc. Jpn.
  {\bf 41}, 513 (1976).

\bibitem{Larkin}  A.I. Larkin, Zh. Eksp. Teor. Fiz. {\bf 105}, 1793 (1994)
  [Sov. Phys. JETP {\bf 78}, 971 (1994)].
  
\bibitem{Ov} Yu. N. Ovchinnikov {\it et al.}, Europhys. Lett.
  {\bf 34}, 645 (1996).
  
\bibitem{Melin02} R. M\'elin {\it et al.}, Eur. Phys. J. B \textbf{26}, 417
  (2002); R. M\'elin, K. Biljakovi\'c and J.C. Lasjaunias,
  Eur. Phys. J. B {\bf 43}, 489 (2005).

\bibitem{revue} S. Brazovskii and T. Nattermann, Adv. in Physics
  {\bf 53}, 177 (2004).

\bibitem{Abe} S. Abe, Physica {\bf 143} B, 85 (1986).

\bibitem{ECRYS02} J.C. Lasjaunias {\it et al.}, J. Phys. IV France {\bf 12}
  (2002) Pr 9-23;
  J.C. Lasjaunias {\it et al.},
  J. Phys. Condens Matter \textbf{14}, 8583 (2002).

\bibitem{Lasjau05} J.C. Lasjaunias, R. M\'elin,
  D. Stare\v{s}ini\'{c}, K. Biljakovi\'{c}, and J. Souletie,
  Phys. Rev. Lett. {\bf 94}, 245701 (2005).

\bibitem{Biljak91} K. Biljakovi\'{c}, J.C. Lasjaunias,
  P. Monceau, and F. Levy,
  Phys. Rev. Lett. \textbf{67}, 1902 (1991).

\bibitem{NMR} E. Barthel {\it et al.}, Europhys. Lett. {\bf 21}, 87
(1993).

\bibitem{JLTP03} S. Sahling, J.C. Lasjaunias,
K. Biljakovi\'c and P. Monceau, J. Low Temp. Phys.
{\bf 133}, 273 (2003).
  
\bibitem{Tua} P.F. Tua and J. Ruvalds, Phys. Rev. B 
  {\bf 32}, 4660 (1985).

\bibitem{Suzumura} Y. Suzumura, T. Saso and H. Fukuyama,
  Jpn. J. App. Phys. {\bf 26}, 589 (1987) Supplement 26-3.

\bibitem{Tutto} I. T\"utt\"o and A. Zawadowski,
  Phys. Rev. Lett. {\bf 60}, 1442 (1988).
  
\bibitem{Maki} K. Maki and A. Virosztek, Phys. Rev. B {\bf 39}, 9640 (1989).

\bibitem{Klemme} B.J. Klemme {\it et al.},
Phys. Rev. Lett. {\bf 75}, 2408 (1995).

\end{thebibliography}
\end{document}